\documentclass[preprint,preprintnumbers,amsmath,amssdoes notymb]{revtex4}
\usepackage{amsmath}
\usepackage{amssymb}
\usepackage{bm}
\usepackage{dcolumn}
\usepackage{graphicx}
\usepackage{color}

\begin{document}

\title{Acceleration of a QM/MM-QMC simulation using GPU}

\author{Yutaka Uejima}
\author{Tomoharu Terashima}
\author{Ryo Maezono}
\address{School of Information Science, Japan Advanced Institute of 
Science and Technology, Asahidai 1-1, Nomi, Ishikawa, 923-1292, Japan}

\date{\today}

\begin{abstract}
We accelerated an {\it ab-initio} molecular QMC calculation
by using GPGPU.
Only the bottle-neck part of the calculation is replaced by CUDA
subroutine and performed on GPU.
The performance on a (single core CPU + GPU)
is compared with that on a (single core CPU with double precision),
getting 23.6 (11.0) times faster calculations in single (double) precision
treatments on GPU.
The energy deviation caused by the single precision treatment
was found to be within the accuracy required in the calculation,
$\sim 10^{-5}$ hartree.
The accelerated computational nodes mounting GPU are
combined to form a hybrid MPI cluster on which we confirmed
the performance linearly scales to the number of nodes.
\end{abstract}

\maketitle

\section{Introduction}
GPGPU (General Purpose computing on Graphical Processing Unit) 
\cite{GPU03,GPGPU}
has attracted recent interests in HPC (High Performance Computing)
to get accelerations in reasonable prices.
Such GPUs with the capability of double precision operations 
get to be available now.
Comfortable environments for developing GPGPU, such as 
CUDA (Compute Unified Device Architecture)\cite{CUDA}, also contribute
recent intensive trend for applying it to scientific applications
with much increased portability.
These include 
computational fluid dynamics,
random number generators, 
financial simulations,
astrophysical simulations,
signal processings, 
molecular dynamics, 
electronic structure calculations,
polymer physics {\it etc.}
Numbers of reports achieving the accerelations
by factors of several tens to hundreds
are found on the web site\cite{CUDA2}.
There has been several attempts using GPGPU applied to
{\it ab-initio} QMC (Quantum Monte Carlo) electronic structure 
calculations \cite{ONO09,ESL09}.
These preceding works shows satisfactory efficiencies of acceleration 
achieved and the possibility of GPGPU challenge in this field.
One of the left problem behind would be how to merge the GPGPU 
with  the conventional stream of the development and
maintenance of large scale scientific codes in general manner.
In pioneering works, GPGPU is sometimes provided in the manner
that a typical algorithm is tested in a small scale bench mark code, 
or some independent 'GPU version' of the code is developed 
by re-writting most of the part of the code in CUDA.
Our next interest is, however, to apply it to materials
simulation programs which are practically used by wider
range of users.
Such programs has been developed 
for over tens of years by many contributors working on
a lot of branches of functionality of the code.
The codes are well designed to be universal to
treat wider range of objects from molecules to solids
as well as modeled systems such as electron gas.
Even for a developer, therefore, it has been not possible to
understand the whole part of the code.
Developing 'Independent GPU versions' seems  
not a practical way to keep harmony with maintenance and
version administration of conventional CPU version of the codes.
In this paper we identified the bottle neck of original CPU version
firstly and then developed CUDA version only on the corresponding
subroutine being tiny part of the whole code.
The main body of the code is written in Fortran90 (F90) and
we combined the CUDA subroutine at object code level.
Users can switch back to the original CPU version of the subroutine
if GPU is not available.

Another different point from preceding studies are
that GPGPU here is devoted to accelerate single core performance,
being possible to coexist with current MPI (Message Passing Interface) 
implementation.
In many QMC codes \cite{NEE10,QMCPACK},
MPI parallelization is used to divide up whole sampling tasks into 
processor cores.
In preceding works GPU is used
so that the parallelized tasks are distributed into
GPU cores instead of CPU cores.
Improved performance was obtained because 
the number of cores in GPU exceeds that in CPU.
We didn't take this way because of the following reasons:
Firstly, in practical codes, the parallelized task
contains much larger processes requiring larger memory 
capacities than in limited-purposed benchmark codes.
We don't expect the task is possible to be put in threads running 
on GPU.
As another reason we point out the fact
that the current CPU-MPI implementation is inherently successful
for QMC because of less frequent communications between
processor nodes.
When the number of cores gets massive it is, nevertheless, 
pointed out the problems such as
the load balancing or other bottle neck arising
{\it etc.}
These problems would similarly occur even when the 
parallel cores are replaced by GPU.
Larger number of dense coupled processor cores in GPU
compared with CPU does not so much matter in our QMC case
because inter-core communication is not the bottle neck.
In this work we kept conventional MPI parallelization over CPU cores.
GPU many-core feature is exploited to speed up each sampling task
which is distributed on each CPU core by MPI, being similar to the idea of 
hybrid parallelization such as Open-MP combined with MPI.

As a proper example we applied GPGPU to a QM/MM 
(Quantum Mechanics / Molecular Mechanics) calculation
called as 'FMO-QMC' calculation\cite{MAE07}.
In this case the bottle neck of single core performance is
identified to the part evaluating electrostatic fields due to given charge
densities.
The field is constructed by large amount of summations in a loop 
being fit to GPU acceleration by its many-core feature,
finally getting 23.6 times faster performance
when we compare the performance on a (single core CPU double 
precision + GPU with single precision) 
with that on a (single core CPU with double precision).
We also confirmed the acceleration can be in harmonic
with that by conventional CPU-MPI parallelization.
As is given in the discussion section later, there would still be more space
to improve the acceleration by combining OpenMP with the present work,
or by using a scheme where the GPU is shared by the MPI processes 
running on the same node.
Here we report a work as a first step towards an efficient acceleration 
of the code by replacing only the 'hotspot' with CUDA-GPU.

\par
The paper is organized as follows.
In \S II we briefly summarize the subjects required here,
such as VMC (Variational Monte Carlo method),
FMO (Fragment molecular method), and GPGPU.
In \S III we describe details how to measure the performance,
namely the system to be evaluated and the coding structures.
Results are shown in \S IV and discussions are given in \S V.

\section{Methodologies}
\subsection{VMC}
\label{VMC}
In {\it ab-initio} calculations
the system to be considered is specified 
by a given hermitian operator $\hat H$ called as
Hamiltonian\cite{PAR94}.
The operator includes information about
positions and valence charge of ions, 
the number of electrons, and the form of the potential
functions in the system.
The fundamental equation at electronic level,
called as many-body Schr\"odinger equation,
takes the form of a partial differential equation with 
the operator $\hat H$ acting on a multivariate function 
$\Psi\left(\vec r_{1},\cdots,\vec r_{N}\right)$, called as many-body
wave function, where $N$ denotes the number of electrons in the system.
The energy of the system, $E$, is obtained as the eigenvalue of the partial differential equation.
The equation has the variational functional \cite{HAM94},
\begin{eqnarray}
\label{eq:vmc_energy}
E &=& \frac{\int \Psi^* \hat{H} \Psi \, d{\vec r_1}\cdots d{\vec r_N}}
{\int \Psi^* \Psi \,
d{\vec r_1}\cdots d{\vec r_N}} 
\nonumber \\
&=& 
\frac{\int |\Psi|^2 \cdot\Psi^{-1} \hat{H} \Psi \, d{\vec r_1}\cdots d{\vec r_N}}
{\int |\Psi|^2 \, d{\vec r_1}\cdots d{\vec r_N}} ,
\end{eqnarray}
which is minimized when the above integral is evaluated
with $\Psi$ being an exact solution of the eigen equation.
For a trial $\Psi$ the functional can be evaluated as an average of the
local energy, $E_L \left(\vec r_1,\cdots,\vec r_N\right) 
= \Psi^{-1} \hat{H} \Psi$ over the probability density distribution 
$p(\vec r_1,\cdots,\vec r_N) = |\Psi|^2/\int |\Psi|^2 \, d{\vec r_1}
\cdots d{\vec r_N}$.
In VMC the average is evaluated by Monte Carlo integration technique
using the Metropolis algorithm to generate 
sample configurations $\left\{\vec R_j\right\}_{j=1}^{r}$
distributed by $p(\vec r_1,\cdots,\vec r_N)=p(\vec R)$, where
$\vec R$ denotes a configuration $(\vec r_1,\cdots,\vec r_N)$
as 
\begin{equation}
E = \sum_{j=1}^{r} {E_L\left(\vec R_j\right)} \ ,
\end{equation}
with $r$ being the order of millions typically.
Trial function $\Psi$ is improved so that the integral is numerically minimized.
Several functional forms for $\Psi$ are possible, amongst which
we took commonly used Slater-Jastrow type wave function \cite{FOU01}.
Since each $E_L \left( \vec R_j \right)$ can be
evaluated independently the summation over $j$ can be 
distributed over processors by MPI with enough high
efficiency\cite{FOU01}.
In this work GPGPU is used to accelerate each
$E_L \left( \vec R_j \right)$ evaluation, not applied to this parallelization.
For VMC we used 'CASINO' program package\cite{NEE10}
with the extended functionality for FMO-QMC\cite{MAE07} 
as described in the next section.

\subsection{FMO method}
\label{FMO method}
FMO (Fragment Molecular Orbital) method, \cite{KIT991,KIT992} 
as a sort of QM/MM method, is devised
to treat larger biomolecules in {\it ab-initio} electronic structure calculations. 
To accommodate in available memory capacities with affordable 
computational cost, the whole system is divided into several sub-systems
called as fragments.
Only within the fragments the electrons are treated fully by
quantum mechanics while the contributions from other fragments are 
replaced into classical electrostatic fields formed by charge densities
of electrons and ions.
While molecular orbital methods (MO) or Density Functional 
Theory (DFT) calculations are commonly used to evaluate sub-systems,
QMC, instead, is expected to be powerful to 
get more reliable estimation of electronic correlations 
which is believed to play important roles in biomolecules.
In the framework, FMO-QMC\cite{MAE07}, 
the additional task to evaluate electrostatic fields at each Monte 
Carlo step causes considerable speed-down by around 
50 times longer CPU time than that of normal QMC with
the same system size.

\begin{figure}[h]
\begin{center}
\includegraphics[width=70mm,angle=0]{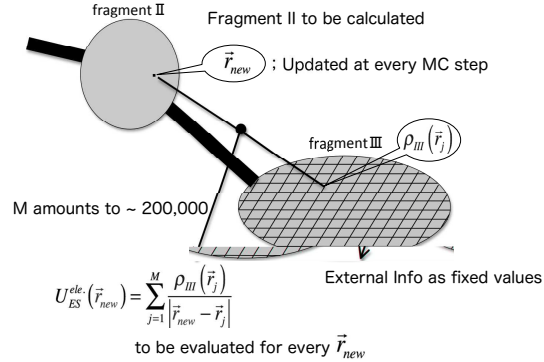}
\end{center}
\caption{Evaluation of electrostatic fields in FMO-QMC calculation.}
\label{cells}
\end{figure}

When we divide the system into $L$ sub-systems,
the energy of the whole system, $E_{\rm All}$, is approximately
evaluated as,
\begin{eqnarray}
E_{\rm All}  
&\approx& \sum_{i=1}^{L-1}\sum_{j=i+1}^{L}E_{ij} 
+ (L-2)\sum_{i=1}^{L}E_i \ ,
\label{FMO}
\end{eqnarray}
from the energies calculated for each sub-system $E_i$,
and those for pairs of sub-systems $E_{ij}$.
These 'fragment energies' are evaluated under 
the electrostatic fields, $U_{\rm ES}\left( {\vec r} \right)$,
due to other fragments.
In FMO-QMC, $U_{\rm ES}\left( {\vec r} \right)$ should be constructed
at every Monte Carlo step with updated electronic positions,
$\vec r = \vec r_{\rm new}$.
Charge densities to form the field are given as input files 
as $\left\{Z_\beta\right\}$
being valence of nuclei and $\left\{\rho\left(\vec r_m\right)\right\}$
being charge intensities of each spatially discretized cell
on the fragment (index $\beta$ runs over $K$ nuclei at
$\vec R_\beta$, and $m$ over $M$ cells centered at $\vec r_m$ 
in the fragment).
The field is hence given as
\begin{eqnarray}
U_{\rm ES}\left( {\vec r_{\rm new}} \right) 
&=& \sum_{m=1}^M \frac{\rho\left({\vec r_m}\right)}
{\left|{\vec r_{\rm new}}-{\vec r_m} \right|}
- \sum_{\beta=1}^K\frac{Z_\beta}{\left|{\vec r_{\rm new}}-{\vec R_\beta} 
\right|}  \nonumber \\
&=& U_{\rm ES}^{\rm ele.}\left( {\vec r_{\rm new}} \right) 
- U_{\rm ES}^{\rm nuc.}\left( {\vec r_{\rm new}} \right) \ .
\label{ElectrostaticField}
\end{eqnarray}
While $K$ amounts to dozens, $M$ gets to around hundreds
thousand, resulting the evaluation of $U_{\rm ES}^{\rm ele.}$
being quite heavy.
Figure \ref{cells} visualizes an image of the evaluation.
The evaluation is the most time consuming part of FMO-QMC,
for which we applied GPGPU acceleration.

\subsection{GPGPU}
GPGPU exploits hundreds of processing cores in GPU
which are originally designed for graphical data processing.
Its performance on single precision operations gets to
tens times faster than that of commonly used CPU.
Comfortable code-developing environments
are available recently, such as CUDA, by which we can
develop GPU codes in more universal manner written in 
language being similar to C language with some extended definitions 
of variables and functions for GPU.
In GPGPU a program consists of host codes
and the kernel codes, former of which run on CPU
while the latter on GPU getting data sent by the host code from CPU.
Frequent data transfer between the host and the kernel
should be avoided
because the transfer is made via bus with relatively low speed.
Less transfers to and more operations on GPU are preferable
for getting better performance.

\begin{figure}[h]
\begin{center}
\includegraphics[width=50mm]{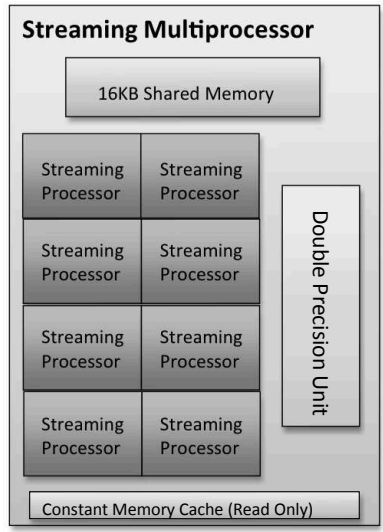}
\end{center}
\caption{Schematic picture of a Streaming Multiprocessor (SM).}
\label{fig:SM}
\end{figure}

In GTX275\cite{GTX275}, a GPU we used here, there are
30 Streaming Multiprocessors (SM).
Each SM includes eight Streaming Processors (SP) which are
used as a smallest processor unit in GPGPU, as shown in 
Fig. \ref{fig:SM}.
Single precision operations can be handled independently on each SP
while double precision requires to be processed
on a DPU (Double Precision Unit) located on each SM.
This makes double precision operations slower by around a factor 
of eight.
Instructions are interpreted on a SM at every four clock cycles
and then executed on eight SPs within the SM.
32 threads (4 cycles $\times$ 8 SP), therefore, forms a
unit of SIMD (Single Instruction Multiple Data) operation,
called as a warp.

In GPU, threads are administrated in a layered structure.
Threads are labeled by three dimensional indices within a block.
Similarly, blocks are labeled by two dimensional indices within a grid,
though the grid is not used in the present study.
Each block is processed by a SM, not by several.
If the number of blocks exceeds that of SM,
the blocks are processed by the SM in due order.
It is therefore usual manner to select the total number of blocks 
to be a multiple of the number of SM.
Since a warp is formed by 32 threads, the total number of
threads would be chosen as a multiple of 32.
From the view point of memory latency it is said 
a multiple of 64 is preferred.
Practically the total number of threads is chosen so that the memory 
capacity required for each thread can be affordable within a SM,
otherwise the performance gets considerably worse.

Table \ref{MemoryList} shows various kinds of memories available in GPU.
The list contains only those relevant to this study, excluding
texture memory.
Off-chip memories are located within GPU board but
not on the device chip.
They have larger capacities and are accessible from hosts
but lower speed in general.
On-chip memories are complementary, namely with higher speed and
lower capacity.
\begin{table}[ht]
\begin{tabular}{c|c|c|c|c|c}
\noalign{\hrule height 1pt}
& Location & Cache & R/W & Availability & Data maintained \\
\noalign{\hrule height 1pt}
Register & On-chip & - & R/W & within a thread & during a thread \\
\hline
Local memory & Off-chip & No &R/W & within a thread & during a thread \\
\hline
Shared memory & On-chip & - & R/W & from all threads & during a block \\
&&&&within a block & \\
\hline
Global memory & Off-chip & No & R/W	& from all hosts & during 
host code\\
&&&& and threads & maintains\\
\hline
Constant memory & Off-chip & Yes & R & from all hosts & during 
host code\\
&&&& and threads & maintains\\
\hline
\end{tabular}
\caption{Various kinds of memory in GPU relevant to this study.
R and W stand for readable and writable, respectively.}
\label{MemoryList}
\end{table}
In GTX275 there are 16,384 registers available for each SM, and
variables defined within kernel codes can be stored there.
When registers are run out, data are evacuated to 
off-chip local memories and newer data are stored into register.
The local memory is about 100 times slower than register
and so it is important to save register for better performance.
Data to be sent to GPU is firstly stored on
a off-chip global memory by a host code 
and then loaded by a on-chip shared memory in usual manner.
Larger capacity is available in global memories ranging from
512MB to 1GB depending on the products.
Again the off-chip global memory is about 100 times slower.
Though they are similarly depicted in Fig. \ref{fig:SM},
the shared memory is on-chip while the constant memory is off-chip.
Each SM has a 16KB shard memory which is accessible from 
all threads within a block.
Though 64KB constant memory is off-chip, it can be accessed with
higher speed from all threads using cache on each SM (constant cache).
This is read only so convenient to store constants defined in kernel codes.
A data load from global memories is executed in parallel manner
by 16 threads simultaneously in GTX275, corresponding
to a half of a warp.
When the addresses accessed by parallel threads
are sequential, the access speed is accelerated by the order of
the number of threads.
This is called as 'coalescing' and very important in the performance
achieved by the present study.

\section{Experimental setup}
\label{Experiments}
As a benchmark system for FMO-QMC, we took a glycine trimer
to measure the performance of GPGPU.
The system is divided into three fragments in this case\cite{MAE07}.
The computational time required to evaluate the energy of the smallest 
fragment ('fr1' in ref \cite{MAE07}),
corresponding to the term $E_1$ included in the second
summation in Eq.(\ref{FMO}), is measured and compared by CPU and GPU.
Detailed setup of the trial wave function such as
basis sets, Jastrow functions, and variational optimizations {\it etc.}
are the same as given in the ref.\cite{MAE07}.
Computational cost for this fragment to 
achieve the statistical error 
required for
meaningful arguments in the context of quantum chemistry, 
as published in reference \cite{MAE07},
is estimated around 50 days with single core,
13,000 times more Monte Carlo steps than the present case.
In this work we took shorten run for benchmark, making it
be finished within around 300 sec. by single core.
Note that the 'accuracy' argued in the present study is different from 
the statistical error because we fixed the seed for the random number
generator, namely we took a deterministic system to be compared
with each other in this work.
\begin{figure}[h]
\begin{center}
\includegraphics[width=90mm,angle=0]{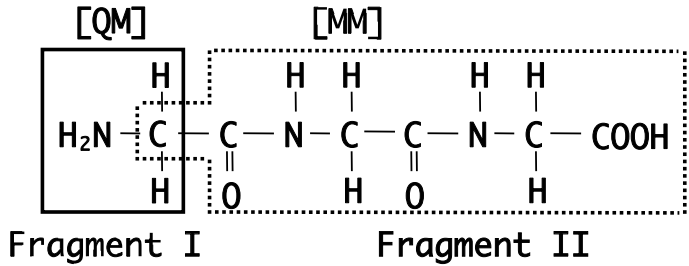}
\end{center}
\caption{Fragmentation of glycine trimer used here\cite{MAE07}.
QM means the fragment treated by quantum mechanical dynamics
while MM the part handled as an environment for QM giving classical 
electrostatic field.
}
\label{fr1}
\end{figure}

\par
\begin{figure}[h]
\begin{center}
\includegraphics[width=90mm,angle=0]{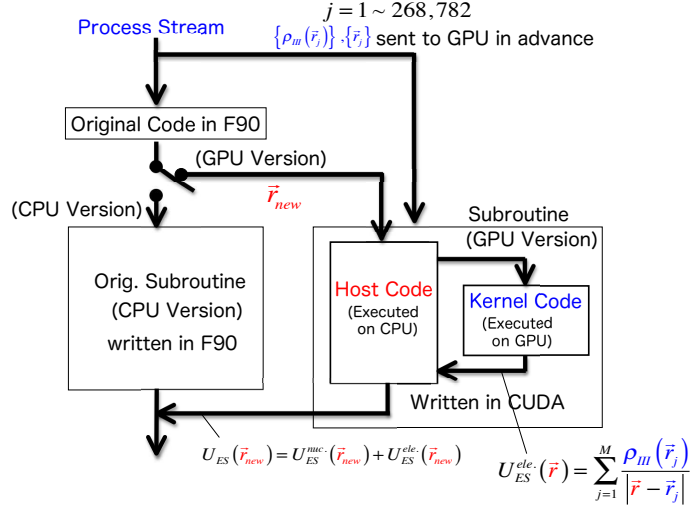}
\end{center}
\caption{Structure of the code.}
\label{codeStructure}
\end{figure}

The FMO-QMC code is an extension of 'CASINO' QMC 
code \cite{NEE10} written in F90, 
while CUDA itself provides only the C-language compiler.
Though there appears commercial fortran compilers for GPU
such as PGI Accelerator Compilers\cite{PGI09} recently,
we didn't take them.
Instead we combined the F90 part and CUDA part at the 
object file level.
The structure of the codes we developed is shown in Fig. \ref{codeStructure}.
We applied GPGPU only to the most time consuming subroutine,
namely that evaluating $U_{\rm ES}\left(\vec r\right)$.
As shown in  Fig. \ref{codeStructure} we developed a detour
leading to GPU version of the subroutine written in CUDA,
consisting of the host and the kernel code.
The host code is called from the main body written in F90,
getting the updated particle position, $r_{\rm new}$,
at each MC step.
The host code then calls the kernel code on which the electrostatic
field, $U_{\rm ES}^{\rm ele.}\left(\vec r_{\rm new}\right)$, is calculated 
to be sent back to the host code.
For more efficiency, the host code calculates 
$U_{\rm ES}^{\rm nuc.}$ independently on CPU, which can be 
finished until it gets $U_{\rm ES}^{\rm ele.}$ from GPU.
These are summed to form $U_{\rm ES}$, which is then sent back
to the main body in F90.
For evaluating $U_{\rm ES}^{\rm ele.}$ on GPU, 
cell positions and charge densities, 
$\left\{\rho\left(\vec r_j\right)\right\}$, should be stored on memories
in GPU.
The data is large but read-only, so the data transfer to GPU is required
only once at the beginning of a run, not consuming computational 
cost relative to the whole CPU time.
The data communication with GPU at each MC step therefore
deals only with $\vec r_{\rm new}$ (input) and $U_{\rm ES}^{\rm ele.}$ 
(output), getting cheaper data-transfer cost.

The summation to form $U_{\rm ES}^{\rm ele.}$ in 
Eq. (\ref{ElectrostaticField}) is divided 
into sub-summations as
\begin{equation}
U_{\rm ES}^{\rm ele.}  = \sum\limits_{j = 1}^M {u_j }  
= B\left[ 1 \right] + B\left[ 2 \right] +  \cdots  + B\left[ {N_B } \right]
\ ,
\end{equation}
and distributed to each block (total $N_B$ blocks) on GPU for acceleration.
Denoting $N_{\rm th}$ being the number of threads within a block,
and $N_{\rm loop}$ being the number of loops per thread,
\begin{equation}
\label{blockSum}
B\left[ m \right] = \sum\limits_{j = 1}^{N_{\rm th}  \times N_{\rm loop} } 
{b_j^{\left( m \right)} } \ ,
\end{equation}
where $\left\{ {b_j^{\left( m \right)} } \right\} \in \left\{ {u_j } \right\}$
are the elements of summation treated by the block $m$.
Total number of terms, $M$ = 268,782, is then distributed to
$\left(N_B \times N_{\rm th}\right)$ threads, within which
$N_{\rm loop}$ loops are performed
so that $M \leq \left(N_B \times N_{\rm th}\times N_{\rm loop}\right)$.
In this work we choose $N_B$=120 as a multiple of the number of SM
(= 30 for GTX275),
$N_{\rm th}$ = 256 as a multiple of warp size, 32,
resulting in $N_{\rm loop}$=9.

$\left\{\rho\left(\vec r_j\right)\right\}$ is initially stored in the global memory.
Getting $\vec r_{\rm new}$ from CPU, it is put in the constant memory,
and then evaluated to form $\left|\vec r_{\rm new}-\vec r_j\right|$, stored
on the register.
Each sub-summation $B\left[m\right]$ is stored in the shared memories
to contribute to the total summation by reduction operation.
Using the read-only constant memory with higher latency for $\vec r_{\rm new}$ 
is found to be essential tip for the present achievement,
because $\vec r_{\rm new}$ is the fixed quantity during the construction
of $U_{\rm ES}$.

Table \ref{exp_specs} summarizes the specification of
a computational node we used for the experiments.
To measure parallel performance of GPU
we used a cluster consisting of four nodes connected by
a 100 Mbps switching hub.
On each node an Intel Core i7 920 processor\cite{INT09} and a GPU
is mounted on a mother board.
Hyper-Threading \cite{HTT} in Core i7 processor is turned off, 
using it as a four-core CPU.		
Specs of GeForce GTX 275\cite{GTX275} is summarized in 
Table \ref{GPUspec}.
Compute Capability specifies the version of hardware level
controlled by CUDA,  above ver.1.3 of which supports double precision
operations.
For Fortran/C codes we used Intel compiler version 10.1.018 for
both using options, '-O3' (optimizations including those for
loop structures and memory accesses), 
'-no-prec-div' and '-no-prec-sqrt' 
(acceleration of division and square root operations with
slightly less precision),
'-funroll-loops' (unrolling of loops), 
'-no-fp-port' (no rounding for float operations),  
'-ip' (interprocedural optimizations across files), and 
'-complex-limited-range' (accerelation for complex variables). 
For CUDA we used nvcc compiler with options
'-O3' and '-arch=sm\_13' (enabling double precision operations).

\begin{table}[ht]
\begin{center}
\begin{tabular}{l|c}
\noalign{\hrule height 1pt}
CPU	 & Intel Core i7 920 2.66 GHz (Max 2.80 GHz)		\\\hline
GPU	 & GeForce GTX 275 $\times$ 1		\\\hline
Motherboard & ASUS RAMPAGE II GENE (Intel X58 chipset) \\\hline
Memory & DDR3-10600 2GB $\times$ 6 \\\hline
OS & Linux Fedora 10	\\\hline
CUDA & CUDA version 2.3	\\\hline
Fortran/C Compiler &  Intel Fortran/C Compiler 10.1.018 \\\hline
MPI &   mpich2-1.2.1 \\\hline
CUDA Compiler & NVIDIA CUDA Compiler (nvcc)  \\\hline
\noalign{\hrule height 1pt}
\end{tabular}
\caption{Setup of a computational node.}
\label{exp_specs}
\end{center}
\end{table}

\begin{table}[ht]
\begin{center}
\begin{tabular}{l|c}
\noalign{\hrule height 1pt}
Compute Capability		&1.3				\\\hline
Global memory & 895 MB			\\\hline
Number of SM &30				\\\hline
Number of SP & 240			\\\hline
Clock of SP &1.404 GHz		\\\hline
Constant memory & 64 KB		\\\hline
Shared memory &16 KB per block 	\\\hline
Warp size & 32				\\\hline
Max number of threads & 512 per block		\\\hline
Memory band width &127 GB per sec.	\\
\noalign{\hrule height 1pt}
\end{tabular}
\caption{Specs of GPU.}
\label{GPUspec}
\end{center}
\end{table}
\clearpage

\section{Results}
Single core performances we measured are tabulated in
Table \ref{Single Core performances}.
The values shown are the CPU time for whole calculation 
including initial data loads onto GPU,
evaluated by averaging over 100 individual runs.
Compared with the normal CPU calculation with double precision 
(341.14 sec.), we finally achieved 23.6 (11.0) times faster 
calculations with single (double) precision by GPU with coalescing.
For more acceleration of the double precision calculation
we also tried replacing our division operation into that
provided as a CUDA function (SFU : Super Function Unit)
but no remarkable speed up observed.
In single (double) precision results the observed deviation
in the final ground state energy from that by the original 
CPU/double precision calculation was within 
$10^{-5}$ ($10^{-12}$) a.u.
This assures the capability of single precision calculations
by GPGPU to provide the results within the chemical accuracy
$\Delta E \sim 10^{-3}$ a.u. with substantially speeding up,
as a particular interest.
\begin{table}[h]
\caption{Comparison of CPU time between single core CPU and GPU.
All values are given in sec.}
\label{Single Core performances}
\begin{center}
\begin{tabular}{c | r | r}
\noalign{\hrule height 1pt}
   & Single Precision & Double Precision\\
\noalign{\hrule height 1pt}
CPU(single core) & - & 341.14 \\\hline
GPU/Coalescing& 14.44 & 30.89\\\hline
GPU/Incoalescing & 44.37 & 48.75\\\hline
\noalign{\hrule height 1pt}
\end{tabular}
\end{center}
\end{table}

As shown in Table \ref{Single Core performances},
the best performance is achieved by the code properly written 
to get coalescing.
The results shown in the row of 'Incoalesing' are obtained
by a naive construction of the summation in Eq. (\ref{blockSum}),
\begin{equation}
B\left[ m \right] 
= \left[ {b_1^{\left( m \right)}  + b_2^{\left( m \right)}  +  \cdots  
+ b_{N_{\rm loop} }^{\left( m \right)} } \right] + \left[ {b_{N_{\rm loop}+ 1}
^{\left( m \right)}  
+  \cdots  + b_{2N_{\rm loop} }^{\left( m \right)} } \right] +  \cdots \\
\ ,
\label{incoalease}
\end{equation}
where $\left[  \cdots  \right]$ corresponds to each sub-summation
evaluated within each thread.
In this construction the threads access to a global memory
to retrieve 
$\left( {b_1^{\left( m \right)} ,b_{N_{\rm loop}  + 1}^{\left( m \right)} ,
b_{2N_{\rm loop}  + 1}^{\left( m \right)} , \cdots } \right)$,
for example at the first step of the loop,
lacking the sequence in addresses to be referred.
By improving the construction as,
\begin{eqnarray}
B\left[ m \right] 
&=& \left[ {b_1^{\left( m \right)}  + b_{1 + N_{\rm th} }^{\left( m \right)}  +  \cdots  
+ b_{1 + \left( {N_{\rm loop}  - 1} \right)N_{\rm th} }^{\left( m \right)} } \right] 
+ \left[ {b_2^{\left( m \right)}  + b_{2 + N_{\rm th} }^{\left( m \right)}  +  \cdots  
+ b_{2 + \left( {N_{\rm loop}  - 1} \right)N_{\rm th} }^{\left( m \right)} } \right] +  \cdots  
\nonumber \\ 
&&+ \left[ {b_{N_{\rm th} }^{\left( m \right)}  +  \cdots  + b_{N_{\rm th}  
+ \left( {N_{\rm loop}  - 1} \right)N_{\rm th} }^{\left( m \right)} } \right] 
\ ,
\label{coalease}
\end{eqnarray}
we can make it to be sequential memory access,
getting coalescing efficiency.
This brought about three times faster evaluation in single precision
calculation.
Without coalescing we could get little acceleration (less than 10$\%$)
in single precision calculation compared with double precision, as seen in 
Table \ref{Single Core performances}.

\begin{figure}[h]
\begin{center}
\includegraphics[width=50mm,angle=-90]{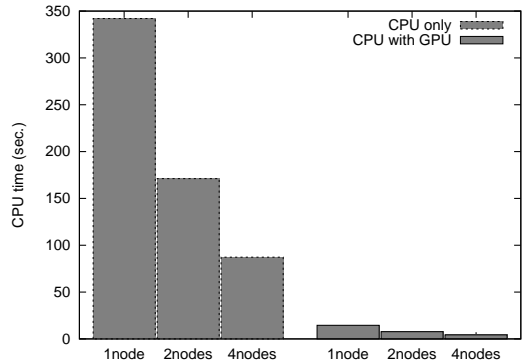}
\end{center}
\caption{Comparison of CPU time between multi-core CPU 
and single precision MPI-GPU.
}
\label{fig:GridBlock}
\end{figure}

\begin{table}[h]
\caption{MPI performances with Double precision. All values are given in sec.}
\label{parallel_1}
\begin{center}
\begin{tabular}{l || c ||c| c}
\noalign{\hrule height 1pt}
   & CPU only & \multicolumn{2}{c}{CPU with GPU} \\
  &  & Double prec(Coalesing). & Single prec. \\
\noalign{\hrule height 1pt}
1 MPI process    &  341.14 (1CPU/1core) & 
30.89 &  
14.44 (1CPU \& 1GPU, 1core/CPU) \\\hline
2 MPI processes & 171.28 (1CPU/2cores) & 
15.70 &
7.68 (2CPU \& 2GPU, 1core/CPU)  \\\hline
4 MPI processes & 87.01 (1CPU/4cores) & 
8.20 & 
4.26 (4CPU \& 4GPU, 1core/CPU) \\\hline
\noalign{\hrule height 1pt}
\end{tabular}
\end{center}
\end{table}

Parallel performances are evaluated and compared with
multi-core CPU, as summarized in Table \ref{parallel_1} and 
Fig. \ref{fig:GridBlock}.
Even the worst case of GPU
(double precision/single core/incoalesing)
(48.75 sec.) 
is still faster than four-core CPU calculation (87.01 sec.).
For 'CPU only' calculations we measured the performance
within a node (and hence the MPI runs within a CPU), 
while for GPU parallel ('CPU with GPU'), $N$-MPI 
runs on $N$ nodes and then only a processor core is
used in a CPU on a node.
Both in 'CPU only' and 'CPU with GPU', measured performances
roughly scale to the number of processor cores, 
showing high parallel efficiency of the Monte Carlo
simulation even with cheaper 100 Mbps switching hub.
The results support that 
the acceleration of single core performance by GPU 
can be in harmony with the MPI parallel acceleration.

\section{Discussions}
Table \ref{performances} shows a rough estimation of
performances expected in devices used here, just by their
numbers of cores and clock frequencies.
\begin{table}[h]
\caption{Estimation of Performances of devises used here.}
\label{performances}
\begin{center}
\begin{tabular}{l || r | r|r}
\noalign{\hrule height 1pt}  
   & Clock Freq. & \# of Cores & Performance \\
\noalign{\hrule height 1pt}  
CPU & 2.66 GHz & 4 cores & 42.56 GFLOPS \\\hline
GPU (Double Prec.) & 1.404 GHz & 30 cores & 84.24 GFLOPS \\\hline
GPU (Single Prec.) & 1.404 GHz & 240 cores & 1010.88 GFLOPS \\
\noalign{\hrule height 1pt}
\end{tabular}
\end{center}
\end{table}
Values to be compared with our achieved factor
23.6 (11.0) for single (double) precision would be evaluated as follows:
Since we got the factors based on CPU single core performance,
42.56/4=10.64 GFLOPS,
we hence expect the upper limit of
the acceleration factor 
being around 94.9
( = 1010.88/10.64) [7.9 ( = 84.24/10.64)] for
single [double] precision operation.

\par
The peak GPU performance for single precision, 
1010.88 GFLOPS,  is simply estimated
as 1.404 GHz $\times$ 240 cores$\times$ 3, where
the last factor, three, is the maximum possible number of operations
at one clock cycle.
Such a peak case occurs only when all the operations
consist of fused multiply-add and a multiply operation,
which fit to the execution by SFU pipeline.
One cannot expect such an extreme case generally and then it is more likely 
being around 337 GFLOPS in practical cases by dropping the last 
factor, three.
Correspondingly the ideal limit of acceleration factor
in the practical situation for single precision
would be evaluated as 31.7 to be compared with our 23.6.
The ideal limits would be achieved when a code all consists of
operations.
Memory accesses contained frequently
in the actual codes would lower the performance,
accounting for the discrepancy.
This would also be supported by the fact that the single precision performance 
strongly depends on the coalescing.
\par
The reduced performance in the double precision
compared with the single precision mainly comes from the fact
that a DPU is available only on each SM, not on SP.
Again, only if the code all consists of double precision operations,
the reduction would occur but in the actual code it wouldn't, 
giving the possibility for acceleration factor being beyond 7.9.
This would account for our achievement with coalescing being 
the factor of 11.0.
The excess factor $7.9/11.0=0.72$ might be attributed to insufficient tuning
on the original CPU code.
If so our achievement in single precision calculation would be reduced as
$23.6\times0.72=17.0$, being still a satisfactory efficiency.
For more reliable/fair estimation of the acceleration factor,
the original CPU version should be optimized enough, though
it is generally difficult to say how much one's code is optimized.
For reference we took a profiling of the code 
using 'OProfile'  profiler \cite{oprofile}.
Measured on Intel Core2Quad/9550, the bottle-neck subroutine
of CPU version (that shown in Fig. \ref{codeStructure}) achieved 0.77 GFlops with 
97.35\% of the whole CPU time.
The value is obtained from the count of operations divided by the
execution time consumed by the subroutine, corresponding
only to less than 2\% of the peak performance of
the processor, 45.28 GFlops\cite{core2}.
It is, however, known that OProfile tends to underestimate 
the performance because it cannot correctly take into account SSE.
For calibration we measure the performance of LINPACK in the same way,
giving 9.4 GFlops by OProfile, while LINPACK itself reports 21.41 GFlops
in its output,
supporting our 'less than 2\%' might be underestimated.
Another possible reason for such low performance would be
because of the dividing operation to get $1/r$ potential.
The peak performance based on multiple/add operation would be
reduced for the dividing operation, and hence corrects the 
measured performance upward.
\par
For more information about how much  the original CPU code optimized,
we examined the dependence on compilers.
Using PGI fortran compiler (ver. 11.1), our best performance 
is obtained with options,
'-03', '-fastsse' (optimization for SSE/SSE2), 
'-tp nehalem-64' (for Intel Core i7(nehalem)),
and '-Mfprelaxed' (accelerating of dividing and squared root operations
with reduced accuracies), getting
343.51 sec. for 1CPU/1core compared with 341.14 sec. by Intel compiler.
This insensitive result is in contrast to the case when we compared them for a
typical example run of CASINO without FMO, namely without running through
the bottle-neck subroutine considered here, showing 
1.62 times faster optimization by intel than that by PGI.
This would also imply that the bottle-neck has enough simple structure 
with little possibility to be optimized further at compiler level.

\par

\par
The acceleration factor by the coalescing is said to be around 10.0 at most.
Though our achievement in total CPU time was only 3.07
as shown in Table \ref{Single Core performances},
our profiler analysis indicates that the execution time consumed
only by the kernel code is accelerated around the factor of 6.0 
by the coalescing with glb\_64b and glb\_128b being increased from zero,
being a satisfactory efficiency.

Reduced/limited performance in double precision calculations
is expected to be improved in next generation GPUs\cite{Fermi09}.
\begin{table}[h]
\caption{Comparison of performances by GTX480 and GTX275}
\label{gtx480}
\begin{center}
\begin{tabular}{l || c| c ||c| c}
\noalign{\hrule height 1pt}
  & \multicolumn{2}{c}{GTX480}  & \multicolumn{2}{c}{GTX275} \\
  &  Double prec. & Single prec. & Double prec. & Single prec. \\
\noalign{\hrule height 1pt}
1 MPI process    &  18.28 &  12.58 &
30.89 &  
14.44  \\\hline
\noalign{\hrule height 1pt}
\end{tabular}
\end{center}
\end{table}
We did a brief check on the dependence of performance on
the generation of architecture using GeForce GTX480 as shown
in Table \ref{gtx480}.
GTX480 is a product employing the latest Fermi architecture\cite{Fermi09} on which
the double precision performance is much improved.
In this quick check we used the same kernel code, not optimized specific for
GTX480.
Because of the available matching to drivers and OS,
the test condition is not the same, using
CUDA version 3.1 and Linux Fedora 12.
Even without further tuning for GTX480
the performance is considerably improved, especially for double precision
being 1.69 times faster.
This comes from the increased number of double precision
operation unit in Fermi.
The number is 16 per SM in GTX480 while one for GTX275.
Having 15 SMs in total, the new architecture has
240 double precision operation units, compared to 30 for 
GTX275.
It is then expected eight times faster performance though,
NVIDIA limits it to be 1/4 of that for this product.
It leads to twice faster performance as expected,
well compared to our achievement, 1.69.
The limitation is removed only for the product line,
Tesla C2000 series, on which more performance is expected.
Another possibility for further improvement
would be to use hybrid parallelization.
During the CPU-GPU operation in the present implementation
only a processor core in CPU is used leaving other three cores unused.
There are still more spaces to increase our efficiency by applying 
OpenMP, for example, to the host code shown in Fig. \ref{codeStructure}
to be exploited unused cores.
\par
Though for practical usage of the application
the code is indeed accelerated by the factor of 23.6,
we point out the statement 'how much the 
GPU accelerates the calculation' includes the ambiguity
which easily leads to misunderstandings especially when
it is argued in the context of architecture performance.
Our achieved factor, 23.6, would be reduced to
be around 2.0 depending on 
the context, as tabulated in Table \ref{represent}:
We first note that our measurement for single precision is not
a 'clearcut' comparison because we compared
[CPU main body (double prec.) + GPU subroutine (single prec.)]
to [CPU main body (double prec.) + CPU subroutine (double prec.)].
More 'natural' choice for the comparison would be to use
original CPU version with single precision.
As excused in \S I, however, it is practically impossible to get such a 
whole single precision version of the original code which is widely used and
developed/maintained in double precision, being the reason why
we took such a setting for the comparison.
Nevertheless, it is worth pointing out that
the single precision performance of original CPU code,
if it were available,  would give more information about 
how much the original code is optimized as the following reason:
If the original code is well optimized to fit to SIMD enough,
the single precision version can give twice faster CPU time at most
because SIMD can accommodate twice operations for single precision
than for double precision.
In such ideal limit we could measure how much the original code has been
optimized by observing how close the CPU time to
the halved value of that by double precision.
If it is not closer it would imply that
not all the operations are fit to SIMD and hence the original 
would have more spaces to be optimized.
If the code is well optimized it might be possible to get less than the halved
because for single precision the cache is more effectively working
with less cache miss.
That for CPU+GPU version would also be reduced a bit
by replacing the CPU part by single prec. version,
but from the fact that the bottle neck is the GPU part, 
we expect its CPU time is not so changed.
Then we estimate a halved value of 23.6,
11.8 as such an extreme limit estimation of
the acceleration factor on the 'natural' definition,
as shown in the third raw of
Table \ref{represent} as the lowest estimate.
However, based on practical experiences, it is quite unlikely to get 
such an ideal situation having halved CPU time of double precision 
by replacing it to single\cite{TOM10}. 
For reference, LINPACK performance measured on Core i7-860 with Intel C Compiler
11.073 showed only 3-4\% increase in FLOPS\cite{TOM10} by replacing double precision
to single.
Taking 4\% as an estimate we also put 22.70 in the third raw of
Table \ref{represent} as the highest estimate.

If we further includes the possibility of CPU to be accelerated
by its multicore into the definition of 'the comparison between 
a GPU and a CPU', the factor should be divided by four,
the number of cores in our case, getting 11.8/4 = 2.95 for single precision
and 11.0/4 = 2.75 for double precision.
If we argue the 'merit factor', namely how much the acceleration
obtained by adding a GPU on a motherboard {\it instead of} an extra
CPU (dual CPU setup), the factor is further divided by two.
In this measure we get 1.38 for double prec. and 1.48 for single prec.
This merit factor would be accompanied by the further note that
adding an extra CPU can achieve the acceleration without the human 
effort of writing the 'Nvidia-specific' version of the subroutine.
In the above context, the ideal limit (1010.88 GFLOPS) and
practical limit (337 GFLOPS) of the GPU performance
are translated into the merit factors of 5.93 - 11.4 and
1.98 - 3.81, respectively.

\begin{table}[h]
\caption{Several possible ways to represent acceleration factor.
SP (DP) stands for single (double) precision, respectively.}
\label{represent}
\begin{center}
\begin{tabular}{c || c | c|c}
\noalign{\hrule height 1pt}  
Reference to estimate & SP on GPU & DP on GPU & Remarks \\
accerelation factor     &                  &                  & \\ 
\noalign{\hrule height 1pt}  
CPU/SingleCore/DP & 23.6 & 11.0 & Practically observed here \\\hline
\parbox[c][0cm][c]{0cm}{}
CPU/SingleCore/SP & 11.8 - 22.70 & N/A & True comparison for SP \\
&&& (ideally estimated) \\\hline
CPU/MultiCore/SP & 2.95 - 5.68 & 2.75 & Comparison between \\
&&&multicore  CPU and GPU \\\hline
CPU plus added  & 1.48 - 2.34 & 1.38 & "Merit factor" \\
CPU/MultiCore/SP  & &  & 5.93 - 11.4 (ideal limit) \\
instead of GPU &&&1.98 - 3.81 (practical limit)\\\hline
\noalign{\hrule height 1pt}
\end{tabular}
\end{center}
\end{table}

System size dependence of the present acceleration should be mentioned.
For the present QM/MM methods (FMO-QMC\cite{MAE07}), the size of MM part matters
for the total CPU cost via the construction of $U_{\rm ES}$.
This is in contrast to SCF (self-consistent field\cite{SZA89}) based methods such as 
FMO-SCF\cite{MAE07}, for which QM size usually matters.
The present system shown in Fig. \ref{fr1} provides the largest MM size among the
fragmentations of the system, and hence
the most expensive CPU time.
The CPU cost scales to the total loop size $M$ which is roughly proportional to
the cube of the MM system size.
QMC calculation itself is known to have such scaling that the CPU time proportional to
$N_{\rm QM}^{2}$ -  $N_{\rm QM}^{3}$, where $N_{\rm QM}$ stands for QM system size
\cite{NEE10}.
In FMO-QMC more than 90\% of the CPU time is spent for the evaluation of MM part,
namely the construction of $U_{\rm ES}$.
Then we expect the total CPU time is almost dominated by MM size.
The present MM size, 19 atoms with 84 electrons, is within the range of usual choice 
commonly used for FMO applied to amino acids, so the results estimated here
give universal trend for other FMO-QMC systems to some extent.
The factor of the acceleration is expected to be unchanged or a bit improved
when the MM size gets larger for the following reasons:
The acceleration is achieved by dividing the total loop size into 
smaller ones each of which processed on parallel threads on GPU.
Such 'barrel processing' gets more advantage as the number of threads 
increased with more efficiency to hide the latency.
The number of variables transferred between CPU and GPU during main calculation,
$\vec r_{\rm new}$ and $U_{\rm ES}$, does not depend on the MM size,
and hence no increase in communication cost.
The capacity to accomodate $\left\{\rho\left(\vec r_{j} \right)\right\}$ increases
but is kept within the range of the global memory which has enough space.
Registers and shared memories are used to accommodate each sub-summation, 
so their capacity limitation does not matter for the choice of MM size.

\section{Concluding Remarks} 
\label{section:concluding_remarks}
We applied GPGPU to accelerate the single core performance 
on a QMC code combined with a QM/MM treatment in FMO method.
Only the bottle-neck subroutine of the code is
translated to be written in CUDA and performed on GPU.
A large scale summation in the part is divided into sub summations
distributed to threads running on many cores in GPU,
getting 23.6 (11.0) times faster performance in single (double) precision
when we compare the performance on a (single core CPU double 
precision + GPU with single precision) 
with that on a (single core CPU with double precision).
The accuracy in single precision calculation was confirmed to be 
kept within the required extent (chemical accuracy, $\sim$0.001 hartree
in energy).
Such accelerated nodes are combined to build a MPI cluster, 
on which we confirmed the MPI performance scaling linearly 
with the number of nodes upto four.
Achieve factors of the acceleration are compared with
ideal limits, and possible accounts for the discrepancy
are investigated,
putting the present work as a first step towards 
further efficient acceleration of such strategy replacing 
only the most time consuming subroutine with CUDA-GPU one.

\section{Acknowledgments}
Authors would like to thank Hisanobu Tomari and 
Daisuke Takahashi for their valuable comments on
CPU single precision performance. 
The computation in this work has been partially performed using
the facilities of the Center for Information Science in JAIST.
Financial support was provided by
Precursory Research for Embryonic Science and Technology, 
Japan Science and Technology Agency (PRESTO-JST), and
by a Grant in Aid for Scientific Research on Innovative Areas 
"Materials Design through Computics: Complex Correlation and 
Non-Equilibrium Dynamics
(No. 22104011)''
(Japanese Ministry of Education,Culture, Sports, Science, and Technology ; KAKENHI-MEXT) for R.M.



\begin{thebibliography}{99}
\bibitem{GPU03}
“ACM Workshop on General Purpose Computing on Graphics Processors”，\\
\newblock http://www.cs.unc.edu/Events/Conferences/GP2/，
\newblock Aug. 2004
\bibitem{GPGPU}
W. Hwu, {\it 'GPU Computing Gems'},
Morgan Kaufmann (2010).
\bibitem{CUDA}
J. Sanders and E. Kandrot, {\it
'CUDA by Example: An Introduction to General-Purpose GPU Programming'
}, Addison-Wesley (2010).
\bibitem{CUDA2}
"CUDA Community Showcase", \\
\newblock http://www.nvidia.com/object
\bibitem{ONO09}
T. Ono and S. Tsuneyuki, unpublished (2009).
\bibitem{ESL09}
K. Esler, J. Kim, D. Ceperley, unpublished (2009).
\bibitem{NEE10}
R. J. Needs，M. D. Towler，N. D. Drummond and P. Lopez Rios, 
\newblock J. Phys. Condensed Matter 22，023201 (2010).
\bibitem{QMCPACK}
QMCPACK Wiki，
\newblock http://cms.mcc.uiuc.edu/qmcpack/index.php/
\bibitem{MAE07}
R. Maezono，H. Watanabe，S. Tanaka，M.D. Towler and R.J. Needs，
J. Phys. Soc. Jpn. {\bf 76}， 064301:1-5 (2007)．
\bibitem{PAR94}
R.G. Parr and W. Yang,{\it
'Density-Functional Theory of Atoms and Molecules'},
Oxford University Press (1994).
\bibitem{HAM94}
B.L. Hammond, W.A. Lester Jr., and P.J. Reynolds, 
{\it Monte Carlo Methods in Ab 
Initio Quantum Chemistry}; World Scientific: Singapore, 1994.
\bibitem{FOU01} W. M. C. Foulkes, L. Mitas, R. J. Needs and
G.~Rajagopal, Rev. Mod. Phys. {\bf 73}, 33 (2001).
\bibitem{KIT991}
K. Kitaura，T. Sawai，T. Asada，T. Nakano and M. Uebayasi，
\newblock “Pair Interaction Molecular Orbital Method: An Approximate Computational Method for Molecular Interactions”，
\newblock Chem. Phys. Lett. {\bf 312}，319-324　(1999).
\bibitem{KIT992}
K. Kitaura，E. Ikeo，T. Asada，T. Nakano and M. Uebayasi，
\newblock  “Fragment Molecular Orbital Method: An Approximate Computational Method for Large Molecules”，
\newblock Chem. Phys. Lett. {\bf 313}，701-706　(1999).
\bibitem{GTX275}
NVIDIA, GeForce GTX 275,
\newblock http://www.nvidia.com/object/product\_geforce\_gtx\_275\_us.html

\bibitem{PGI09}
PGI Accelerator Compilers，
\newblock http://www.pgroup.com/resources/accel.htm

\bibitem{INT09}
Intel Core i7-920 Processor,
\newblock http://ark.intel.com/Product.aspx?id=37147

\bibitem{HTT}
Intel Hyper-Threading Technology，\\
\newblock http://www.intel.com/jp/technology/platform-technology/hyper-threading/index.htm

\bibitem{oprofile}
OProfile,
\newblock http://oprofile.sourceforge.net/

\bibitem{core2}
Intel Core2Quad Processor,
\newblock http://www.intel.com/support/processors/sb/cs-023143.htm

\bibitem{Fermi09}
NVIDIA Fermi，
\newblock http://www.nvidia.com/fermi

\bibitem{TOM10}
H. Tomari, private communication.

\bibitem{SZA89}
A. Szabo and N. Ostlund, 
"{\it Modern quantum chemistry-introduction to advanced electronic
structure theory"}, 
McGraw-Hill, New York, 1989.

\end{thebibliography}
\end{document}